# Made in the Image of the Creator?

# AI Use in Creative Work Reduces Authenticity, Love & Other Psychological Benefits

Jessica Carson, Paul H. P. Hanel, and Wijnand A. P. Van Tilburg

## Author Note

Jessica Carson https://orcid.org/0009-0000-1410-5020

Paul Hanel https://orcid.org/0000-0002-3225-1395

Wijnand Van Tilburg https://orcid.org/0000-0002-9724-0603

We have no conflicts of interest to disclose. This research did not receive any specific grant from funding agencies in the public, commercial, or not-for-profit sectors. No portion of this work was generated by AI; the authors used Claude for the purposes of checking spelling, grammar, formatting, and references. Anonymized data, supplemental materials, and Study 2's pre-registration are available on the Open Science Framework (OSF): https://doi.org/10.17605/OSF.IO/8CYB5

**Jessica Carson:** Conceptualization, Methodology, Validation, Formal Analysis, Investigation, Data Curation, Writing - Original Draft, Writing - Review & Editing, Visualization, Project administration. **Paul Hanel** and **Wijnand Van Tilburg**: Methodology, Validation, Formal Analysis, Writing - Review & Editing, Supervision

Correspondence concerning this article should be addressed to Jessica Carson (j.carson@essex.ac.uk) at the Department of Psychology, University of Essex, Wivenhoe Park, Colchester CO4 3SQ, United Kingdom.

**Abstract**

Across psychological, philosophical, and theological traditions, creations have been understood as being "made in the image of their creator"—they reflect the person who made them. The rise of generative artificial intelligence, however, has blurred the throughline between creator and creation. Across three experiments (one pre-registered, U.S. and U.K. samples; $N = 651$), we examined differences in creator outcomes associated with AI-assisted versus human-only creative work. Compared with human-only work, AI-assisted work was associated with significantly lower self- and identity-relevant indicators, including authenticity, self-creation connection, positive affect, and intrinsic motivation, and with higher levels of the "existential ick" emotions of disgust, guilt, and shame. Additionally, we found that the authenticity gap was statistically mediated by the creators' perception that their AI-assisted work contains less of their love, above and beyond differences in perceived effort, uniqueness, quality, and expensiveness. We also ruled out effort justification as an alternative explanation. Together, these findings suggest that AI reshapes not only how creative work is produced, but also the extent to which creators experience their work as an authentic expression and extension of themselves.

*Keywords:* Creativity, Artificial Intelligence, Generative AI, Authenticity, Love, Affect, Motivation

**Made in the Image of the Creator?**

**AI Use in Creative Work Reduces Authenticity, Love & Other Psychological Benefits**

## 1. Introduction

At this point, many creators have sat down with a generative AI model to co-create something—an act researchers have cleverly termed "generative synesthesia" (Zhou & Lee, 2024)—if only out of curiosity. For those who have, two simultaneous realizations may have emerged. First, that AI is an incredibly powerful, if deeply imperfect, tool that can make the creative act temptingly efficient and, at times, quite literally more creative (Zhou & Lee, 2024; Doshi & Hauser, 2024; Jia et al., 2024). Second—and often far less encouraging—that it feels existentially threatening to the very premise of human creativity (Millet et al., 2023). Creators and scholars alike have observed that AI does not lend the same authentic signature of humanity that has long defined the creative act (Runco, 2025; Kharkhurin & Corazza, 2026)—and may therefore not confer the same benefits. Indeed, AI use arouses a host of niggling sentiments in creators: a moral discomfort, including increases in self-conscious emotions like guilt (Chan, 2024; Hass et al., 2026; Cipriani et al., 2026); compromised self-efficacy, ownership, and agency (Lee et al., 2026; He, 2026; Draxler et al., 2024); and concerns about preserving one's authentic voice (Hwang et al., 2025). Such findings suggest that the self-expressive and self-constitutive properties of creativity may be compromised when an AI model becomes an intermediary between creator and creation, and that self- and identity-related outcomes may be adversely impacted as a result.

Our present research extends a growing body of work on audience responses to AI-generated creativity by focusing instead on the internal experience of creators themselves. Although several studies have examined audience perceptions of creator or creation authenticity in relation to AI use (Kirk & Givi, 2025; Raj et al., 2026; Brüns & Meißner, 2024; Koning & Voorveld, 2025; Benegal et al., 2026; Sahebi et al., 2026), no quantitative study, to our knowledge, has examined creators' felt authenticity in relation to their own AI use, much less alongside other self- and identity-relevant outcomes. This has important implications for the psychological well-being of individuals engaged in creative work—from students to artists to professionals—as industries and institutions race to embrace AI tools (Corazza et al., 2025). We hypothesized that, relative to human-only creative work, AI-assisted creativity inhibits a creator's expression of self, and will thus be associated with drops in authenticity, self-creation connection, positive affect, and intrinsic motivation, and a rise in negative affect, specifically the "existential ick" emotions, such as guilt, shame, and disgust.[1] We also hypothesized that perceived love will mediate the authenticity gap, such that creators will perceive their AI-assisted work as containing less of their love, which will in turn diminish feelings of authenticity.

## 2. Theoretical Perspectives on the Self & Creativity

Across religious, philosophical, and psychological thought, creations have long been viewed as extensions of the creators who made them. Perhaps the most famous instance of this arises from the Judeo-Christian doctrine of *Imago Dei*, or "Image of God," according to which humans were created in the likeness of their Creator. An adaptation of this idea—that creators likewise make creations in their own image—can be found throughout theoretical and empirical literature alike. William James (1890) argued that the self is not confined to the mind or body, but extends outward to include the things one owns and makes, a phenomenon Russell Belk (1988) later termed the *extended self*. Carl Rogers (1954) observed that "creativity always has the stamp of the individual upon its product" (p. 250), while Abraham Maslow (1963) proposed that creativity and self-actualization "may turn out to be the same thing" (p. 4). More recently, researchers have found that creative works are liked and valued to the degree that they are perceived to be imbued with the creator's "essence" (Newman & Bloom, 2012; Kreuzbauer et al., 2015; Newman et al., 2014; Smith et al., 2016), and, more specifically, their love (Fuchs et al., 2015). Creators also possess individual creative styles which can be detected across domains of their creative work (Gabora et al., 2012). Research on authenticity similarly finds that creators vary in the extent to which they feel they have expressed themselves authentically (Carson, Smith, et al., forthcoming), and that audiences agree on which creators are being authentic—that is, "true to who they really are" (Henderson & Gabora, 2013, p. 2526).

[1] Guilt and shame were considered in Study 1; disgust was added in Study 2.

This conveyance of a creator, however, is not inviolable. Since the industrial revolution, artists and intellectuals have warned that machines strip creative works of their essence, love, or authenticity: John Ruskin (1853) condemned industrial manufacturing for negating humanity from the act of creation and reducing the creator to a tool, and Walter Benjamin (1935) famously argued that mechanical reproduction withers a work's *aura*—its unique presence. Such critiques, and ensuing preferences, still hold in the modern day. Audiences find handmade works more attractive than machine-made ones (Fuchs et al., 2015), and prefer handmade objects even when their functionality is inferior to those made by machines (Kreuzbauer et al., 2015). This preference for human-made creative works is strongest for products with higher symbolic value (e.g., when expressing something about oneself is of greater importance; Granulo et al., 2021), which has been attributed to the perception that symbolic objects represent a creator's unique expression, materialized in the moment of creation (Kreuzbauer et al., 2015; Gell, 1998). In this sense, creative works are a way to "look into and read" the creator's mind (Humphrey, 2007, p. 747). Human-made creations are therefore believed to reflect, represent, and, even if only symbolically, *contain* part of the creator who made them; conversely, automation and digitalization are believed to standardize the creative work and "wipe out" the unique essence that the creator might have imbued into it (Kreuzbauer, 2016; Kreuzbauer & Keller, 2017).

The AI revolution has resurrected many of the concerns raised during the industrial revolution. The sense that creativity is being appropriated by artificiality, with human creativity becoming increasingly indistinguishable from AI, has unsettled humanity's ontological security—our confidence in our unique position in the world—largely due to the rattling of our anthropocentric creativity beliefs, defined as "the conviction that creativity is a uniquely human characteristic that constitutes the essence of being human" (Millet et al., 2023, p. 3). This angst is abundantly evident in the public's reaction to creative AI use: while humans are generally not very good at distinguishing AI- from human-generated creativity (Porter & Machery, 2024; Köbis & Mossink, 2021; Hitsuwari et al., 2023; Gangadharbatla, 2022), their positive feelings about a work (Raj et al., 2023, 2026; Horton et al., 2023; Millet et al., 2023; Bellaiche et al., 2023; Sahebi et al., 2026) and its creator (Brüns & Meißner, 2024; Koning & Voorveld, 2025; Schilke & Reimann, 2026; Benegal et al., 2026) sour when artificial origins are suspected or revealed. These effects are most pronounced among people with stronger anthropocentric creativity beliefs (Millet et al., 2023), and are mediated by factors like moral disgust and perceived inauthenticity (Raj et al., 2023, 2026; Kirk & Givi, 2025). In parallel, creators themselves report curiously anhedonic effects when using AI to create: they find it less enjoyable and valuable (Mei et al., 2025), experience reduced self-efficacy, meaning (Lee et al., 2026), and ownership (Lee et al., 2026; Draxler et al., 2024; Xu et al., 2024), and face larger subsequent declines in intrinsic motivation (Wu et al., 2025).

Taken together, creators and consumers alike appear to feel a lack of authentic human presence in creative work made with machines—and in particular, AI. And, perhaps resultingly, the positive outcomes that have long been associated with the production and consumption of creative work appear to attenuate, and sometimes reverse, for work co-created with AI. These perspectives provide the basis for examining how AI use may affect self- and identity-relevant dimensions of a creator's experience.

## 2.1 Authenticity, Self-Concept & Identity

**H1:** State authenticity will be lower for AI-assisted creative work than for human-only creative work.

**H2:** Self-creation connection will be weaker for AI-assisted creative work than for human-only creative work.

For many years, researchers have used the Standard Definition of Creativity (SDC) to empirically pin down the study of creativity (Stein, 1953; Runco & Jaeger, 2012; Runco, 2018). According to the SDC, creativity is that which is original (also: novel, surprising, etc.) and effective (also: useful, appropriate, etc.). With the advent of AI, however, the SDC feels hardly satisfactory, even to its originators (Runco, 2023, 2025), as today's models are capable of satisfying both of these criteria (Haase & Hanel, 2023; Holzner et al., 2025). This has led to a flurry of activity within the field to redefine the contours of creativity, such that human creativity can be meaningfully distinguished from AI-generated creativity—what is sometimes called artificial creativity (Runco, 2025). While AI is capable of achieving originality and effectiveness at levels comparable to humans, researchers have observed that AI-generated creativity lacks the distinctive presence associated with human creativity: the rootedness in lived experience and autobiographical memory, intrinsic values and beliefs, personal identity and selfhood, felt emotions and vulnerability, and the creator's unique relationship to their work (Kharkhurin & Corazza, 2026; Corazza et al., 2025; Runco, 2025). In other words, while AI can be novel and effective, it lacks *authenticity*. For this reason, researchers have not only proposed that authenticity is the principal divergence between human and AI-generated creativity (Kharkhurin & Corazza, 2026), but that it should be added as a criterion to the SDC (Runco, 2025).

Authenticity in this context refers to an alignment between the various facets of a creator's individuality—their values, identity, lived experience, preferences, emotions, and imperfections—and their expression (Carson, Smith, et al., forthcoming; Kharkhurin, 2014; Kharkhurin & Corazza, 2026; Runco, 2023). When expressed creatively, this individuality lends a work its nuance and distinctive style. In a typical, human-led creative process, authenticity is a natural outcropping of the individual expressing their inner self and relating their values and beliefs to the world (Kharkhurin, 2014). In AI-mediated creative work, however, the model acts as an intermediary between creator and creation, which may threaten the direct correspondence that allows a work to be authentic in the first place. The AI model offers opinions, corrections, critiques, additions, negations, and other alterations that do not arise organically from the creator, and thus acts as an external influence on the creator and their work, which is typically considered adversarial to authenticity (Wood et al., 2008; Kernis & Goldman, 2006). The authenticity of a work may therefore be considered compromised across all 4 Ps of creativity (Rhodes, 1961): the *process*, as the creator prompts the model, evaluates its response, and adapts the work accordingly; the *person*, as the creator works with an artificial collaborator whose outputs reflect ideas and influences beyond their own; the *press*, as the creative act occurs within the constraints and affordances of a digital tool; and, of course, the *product*, as the work contains some degree of synthetically generated material and will ultimately diverge, subtly or substantially, from what the creator would have produced on their own.

Indeed, when creators share their greatest concerns around co-creating with AI—which include credit, authorship, ownership, control, and agency—researchers note that these trace back to concerns about preserving their authentic voice (Hwang et al., 2025). It is possible that these reductions in authenticity arise from self-deterministic threats to autonomy (Ryan & Deci, 2004), and from the reliance on external regulation (Goldman & Kernis, 2002) by an artificial system. Supporting this, passive or significant AI use for creative tasks has been found to significantly undermine feelings of self-efficacy, ownership, meaningfulness, and agency in creators (Lee et al., 2026; He, 2026; Draxler et al., 2024), all of which are strongly tied to feelings of authenticity (Ryan & Deci, 2004; Goldman & Kernis, 2002). Revealingly, creative expertise moderates some of these effects, such that experts experience more severe declines in ownership and agency than novices, likely because AI displaces the skills and judgments on which experts' identities are formed (He, 2026).

But creativity does more than express an authentic, already-formed self. It also allows creators to construct the identity and define the self (Barbot & Heuser, 2017; Sica et al., 2019; Culpepper & Gauntlett, 2024), counteract identity loss and uncover new aspects of the self following adversity (Reynolds & Prior, 2006; Reynolds & Vivat, 2010; Forgeard, 2013), and organize complex emotional experiences into a self-narrative (Pennebaker & Seagal, 1999). Creative works can therefore serve as both personal and social representations of self-discovery, creative accomplishment, and one's unique identity (Culpepper & Gauntlett, 2024; Glăveanu & Tanggaard, 2014). In AI-assisted work, however, it is possible that these self-verifying and self-enhancing benefits attenuate. When elements of a work originate from an AI model, the creator may feel as though the resultant work is less representative of themselves, which may therefore compromise their willingness to incorporate it into their self-concept. This may be related to less direct engagement on behalf of the creator's self (Kharkhurin & Corazza, 2026), greater guilt associated with outsourcing a personally meaningful act (Hass et al., 2026), or diminished self-efficacy, ownership, meaningfulness, and agency (Draxler et al., 2024; He, 2026; Lee et al., 2026). To assess the extent to which creators use a creation to construct the self, create their personal identity, and communicate the self-concept to others, we introduce the concept of self-creation connection. We predict that, relative to human-only work, creators will experience AI-assisted creations as less compelling vehicles to form the self and express it to others.

**2.2 Motivation & Affect**

**H3:** Intrinsic motivation will be lower for AI-assisted creative work than for human-only creative work.

**H4:** Negative affect will be higher for AI-assisted creative work than for human-only creative work, specifically "existential ick" emotions like disgust, guilt, and shame.

**H5:** Positive affect will be lower for AI-assisted creative work than for human-only creative work.

Creativity has long been understood as an intrinsically rewarding act, one pursued in the name of curiosity, enjoyment, learning, and growth, rather than for external reward (Amabile, 1983, 1996; Ryan & Deci, 2000). Csikszentmihalyi (1975, 1990) frequently described the creative act as *autotelic*, meaning it is purposeful in and not apart from itself. This intrinsic satisfaction arises from the creator's active participation in which they exercise their capacities, overcome difficulties, express their identities, and actualize their potentialities (Maslow, 1963; Rogers, 1954; May, 1975). Intrinsic motivation is therefore associated with a range of psychological benefits, including positive affect (Amabile, 1996; Ryan & Deci, 2000; Stanko-Kaczmarek, 2012). Indeed, creators who are more intrinsically motivated experience significantly higher levels of positive affect during their creative process, and rate their performance significantly higher (Stanko-Kaczmarek, 2012). But evidence suggests that AI use compromises this motivational profile. As discussed in the section prior, AI use diminishes feelings of autonomous engagement, like self-efficacy, ownership, meaningfulness, and agency (Lee et al., 2026;

He, 2026; Draxler et al., 2024). During AI-assisted creativity, creators may become less self-directed and more responsive to the model as it generates ideas, resolves difficulties, or steers the direction of the work. Even if this collaboration results in a work which is technically more creative, it may still reduce the feelings of competence and autonomy that are typically associated with creativity. Consistent with this, creators who use AI show larger subsequent declines in intrinsic motivation and greater increases in boredom than creators who rely on their own effort (Wu et al., 2025). We therefore predict that creators will report lower intrinsic motivation when reflecting on AI-assisted than on human-only creative work. Further, because intrinsic engagement is so entangled with positive affect, we predict that AI-assistance will also be associated with lower positive affect.

Beyond declines in positive affect, there is emerging evidence that AI use may increase negative affect as well. While quantitative research on creators' emotional experiences remains limited, Hass et al. (2026) found that individuals felt greater guilt when using AI to write emotionally meaningful messages than when writing those messages themselves, an effect explained by the feeling that passing off AI-generated work as one's own is dishonest. The recently introduced shame and guilt related to AI tools scale, or SAG-AI (Cipriani et al., 2026), which was published after our data collection, lends further credence to the idea that moral emotions—the sense one has been dishonest or deceptive, or has behaved in a transgressive or unethical way—are inflamed during AI use. Research on audience reactions provides strong though indirect evidence that AI-mediated creativity arouses negative affect. Creative works associated with AI, or creators suspected of using AI, are met with a host of negative sentiments, from disgust to distrust to feelings of uncanniness (Kirk & Givi, 2025; Raj et al., 2026; Amos & Zhang, 2024; Rapp et al., 2025). Although these studies explore the perception of audiences rather than creators, they suggest that the use of AI triggers moral concerns around sincerity, authorship, effort, and authenticity. We therefore hypothesize that AI-assisted creations will be associated with increased negative affect, specifically the moral, "existential ick" emotions.

**2.3 Love & Other Considerations**

**H6a-e:** We expect that AI-assisted creative work will be perceived as less unique, less effortful, lower in quality, less expensive, and containing less love than human-only creative work.

**H7:** The perception that the creative work symbolically contains love will mediate the effect of AI use on authenticity, even when perceived uniqueness, effort, quality, and expensiveness are included as alternative mediators.

When it comes to evaluating creative works, humans engage in two related forms of magical thinking. The first is psychological essentialism, the cross-cultural tendency, observed in children as young as three or four, to believe that people and things possess deep, hidden, immutable properties, or essences, that define their true nature (Medin & Ortony, 1989; Gelman, 2003; Newman, 2016). The second is contagion beliefs: the idea that objects, like creative works, can take on a person's essence through physical or symbolic contact with them (Johnson & Jacobs, 2001; Nemeroff & Rozin, 1994; Frazier et al., 2009; Newman & Dhar, 2014). Together, these tendencies lead people to believe that creative works become contaminated with the essence of their creator (Newman & Bloom, 2012, 2014; Newman, 2016; Newman et al., 2014). Further, when a creator has an essence that is valued, the degree of this contamination enhances perceptions of liking and value. This helps explain why consumers pay more for creative works that possess greater physical proximity to their creator (Newman & Bloom, 2014) or greater temporal proximity to the moment of creation, as in the case for works with lower serial numbers (Smith et al., 2016). Conversely, it explains why consumers pay less for objects that have interacted with a desirable essence, but have been sanitized or sterilized prior to receipt (Newman & Bloom, 2014; Newman et al., 2011). People also believe the essence contained within an object can transfer to them and bestow them with the special gifts of its owner or creator, a belief that can produce actual increases in confidence and performance (Lee et al., 2011; Kramer & Block, 2014).

One aspect of a creator that is believed to be imbued into their work is their love. Most people have experienced this at some point: they are partial to a sweater handknit by a grandmother or local craftsperson over a near-identical one from a big-box store, not because they differ tremendously in their degree of uniqueness, expensiveness, quality, or effort, but because the former is felt to be imbued with the creator's love. Fuchs et al. (2015) refer to this as the *Handmade Effect*. Audiences find handmade works more attractive than machine-made ones, an effect statistically mediated by the belief that the former contain the artisan's love, even when perceived uniqueness, expensiveness, quality, and effort are accounted for. They define love as the creator's "warmhearted passion for a product or its production process" (p. 100), which is perceived by consumers as an "emotion of strong attraction and passionate attachment" (p. 99) that becomes symbolically imbued into the work. The fact that the love effect persists after accounting for effort is meaningful, as researchers have long theorized that people value things more specifically because greater effort has been put into them (Aronson & Mills, 1959; Kruger et al., 2004; Norton et al., 2012). It is apt today to ask whether the same effect applies to human-only versus AI-assisted creations, and extends to the creator's perception of their own work. We predict this to be the case. We further predict that the difference in love will mediate the authenticity gap between AI-assisted and human-only creative work. Such a finding would indicate that love is a mechanism by which a creative work comes to feel—or fails to feel—made in its creator's image.

## 3. The Present Studies

Across three studies, we tested whether AI-assisted creative work disrupts the sense that a creation is "made in the image of its creator." In Study 1, we used a within-subjects design to compare creators' felt authenticity, self-creation connection, positive and negative affect, intrinsic motivation, and, exploratorily, self-assessed creativity for a creative work they made on their own versus with AI-assistance. In Study 2 (pre-registered), we sought to replicate these findings, and test our hypothesis that symbolic love mediates the authenticity gap between AI-assisted and human-only work, above and beyond four rival mediators. In Study 3, we aimed to replicate the mediation, as well as test whether love can be experimentally manipulated, and if this produces expected changes in authenticity. Together, these studies offer the first empirical account of how and why AI-assistance compromises the self-constructive and self-expressive properties of creativity. All studies were reviewed and approved by our local ethics committee (S1: ETH2526-0941, April 7, 2026; S2: ETH2526-1149, May 29, 2026; S3: ETH2526-1452, July 10, 2026).

## 4. Study 1

In Study 1, we examined creators' feelings of authenticity while producing AI-assisted versus human-only creative work using a within-subjects design. We also tested for differences in self-creation connection, positive and negative affect, intrinsic motivation, and self-assessed creativity.

### 4.1 Method

#### *4.1.1 Participants*

A power analysis revealed that to detect an effect of $d$ = 0.30 (Fuchs et al., 2015) with a power of .95, 147 participants were needed. Two hundred one U.S. adults ($M_{age}$ = 37.4 years, $SD$ = 12.7, range: 18–78; 99 male, 101 female, 1 non-binary/third gender) were recruited via Prolific and compensated at minimum wage. Participants were eligible if they were 18 years or older, fluent in English, and U.S. citizens living in the United States who used AI at least 2 times per week. Informed consent for all studies was gathered electronically.

#### *4.1.2 Procedure*

After providing informed consent, participants completed two creative work reflection tasks in randomized order. They were asked to reflect on a creative work spanning any domain that they created on their own, or with the assistance of AI (prompts in supplemental materials). After each reflection task, participants completed scales with respect to the creative work they had just described: the state authenticity scale, self-creation connection scale, the positive and negative affect schedule, and motivation scales. We also asked for descriptive information about the creation (e.g., category of creative work, extent of AI use, etc.). Demographic information was collected at the end of the study.

#### *4.1.3 Measures*

All scale items, anchor points, and prompts are provided in the supplemental materials. Unless otherwise noted, items were rated on seven-point scales (1 = strongly disagree; 7 = strongly agree).

*Authenticity.* State authenticity—the experience of feeling authentic while creating each work—was measured with the 6-item state authenticity scale (Horton et al., 2025), adapted to past tense to reference the creation experience (e.g., "I feel authentic" → "I felt authentic"). The scale includes three authenticity items (e.g., "I felt like myself") and three inauthenticity items (e.g., "I felt like I was wearing a 'false-face'"). The inauthenticity items were reverse-scored and all six items were averaged into a single authenticity score ($\alpha_{AI}$ = .91, $\alpha_{human}$ = .83).

*Self-creation connection.* The extent to which participants use a specific creation to construct their self-concepts and personal identity was measured with seven items adapted from the self-brand connection scale (Escalas & Bettman, 2003), with the referent changed from a brand to the creative work (e.g., "Brand X reflects who I am" → "This creation reflects who I am," $\alpha_{AI}$ = .97, $\alpha_{human}$ = .93).

*Affect.* Participants reported the extent to which they experienced various emotions while producing each creative work using the positive and negative affect schedule (PANAS; Watson et al., 1988). Positive affect (10 items; e.g., "inspired," "proud"; $\alpha_{AI}$ = .96, $\alpha_{human}$ = .90) and negative affect (10 items; e.g., "guilty," "ashamed"; $\alpha_{AI}$ = .95, $\alpha_{human}$ = .94) scores were formed by averaging the items. Items were rated on a seven-point scale (1 = very slightly or not at all; 4 = moderately; 7 = extremely).

*Motivation.* Participants' motivations for creating each work were measured with 12 items developed for this research. Six items captured intrinsic motivation (e.g., "To grow and learn new things," "To express my true self"; $\alpha_{AI}$ = .88, $\alpha_{human}$ = .82) and six items captured extrinsic motivation ("To earn money," "To gain fame or visibility"; $\alpha_{AI}$ = .90, $\alpha_{human}$ = .91). Items were rated on a seven-point scale (1 = not at all; 4 = moderately; 7 = extremely).

*Self-assessed creativity.* A single item captured how creative the participants felt their work to be (1 = not creative at all, 7 = extremely creative). This approach is in keeping with prior research on participants' perception of the creativity of their own work (Silvia et al., 2017; Gable et al., 2019; Beaty & Johnson, 2021).

### *4.2 Results and Discussion*

*Authenticity.* As predicted, participants reported significantly lower authenticity for their AI-assisted than human-only creative work (Table 1). This gap suggests that AI use in the creative process interrupts "the subjective psychological state of feeling like one's self — unburdened by constraint, inhibition, or pretense" (Horton et al., 2025, pp. 2-3). We suggest this is because an intermediary—in this case, the AI model—creates, at least in part, on behalf of the creator, preventing the full expression of self in the work.

*Self-creation connection.* As predicted, self-creation connection was significantly lower for AI-assisted than human-only work. This finding suggests that creators feel less able to "construct the self or communicate the self-concept to others" (Escalas & Bettman, 2003, p. 3) through AI-assisted work.

*Motivation.* As predicted, intrinsic motivation was significantly lower for AI-assisted work. While we did not have a hypothesis with regard to extrinsic motivation, exploratory analyses revealed that extrinsic motivation was also significantly lower for AI-assisted work, although this effect was descriptively less than half the size of the intrinsic motivation effect. This may reflect that extrinsic motivation is tied more to the output, while intrinsic motivation is tied more to the process; when using AI, the process may feel significantly less driven by the creator, leading to a larger drop in intrinsic motivation, while the output is only partially compromised.

*Affect.* Positive affect was significantly lower in the AI-assisted condition, while negative affect did not significantly increase. However, exploratory item-level analyses revealed the only highly significant negative affect items ($p < .001$) were guilt and shame, which supports our "existential ick" hypothesis and suggests a more refined measure of moral emotions may be useful. We interpret this as a kind of "creative anhedonia:" the positive emotions typically associated with creativity (e.g., interest, determination, inspiration) flatten out during AI-assisted work, while negative affect only rises for self-conscious moral emotions. This may be related to our observed drop in intrinsic motivation, which is widely understood as concomitant with both creativity and well-being; indeed, positive affect and intrinsic motivation were strongly correlated in both the human-only ($r = .59$, $p < .001$) and AI-assisted ($r = .79$, $p < .001$) conditions.

*Self-assessed creativity.* An exploratory finding, participants rated their AI-assisted work as significantly less creative than their human-only work. This mirrors, in reverse, research showing that intrinsically motivated creators experience higher levels of positive affect and rate their performance more favorably, via a dynamic process whereby intrinsic motivation generates more positive affect which generates more favorable views of their work  (Stanko-Kaczmarek, 2012). Given that AI use appears to interrupt this intrinsic engagement, it follows that downstream benefits—like feeling good about the process and evaluating one's work favorably—may be similarly compromised.

**Table 1**

*Study 1: Descriptive Statistics, Paired t Tests, and Effect Sizes for All Measures*

| | Human-only *M* (*SD*) | AI-assisted *M* (*SD*) | *t* | *df* | *p* | *dz* |
|---|---|---|---|---|---|---|
| Authenticity | 6.14 (1.00) | 4.95 (1.49) | 9.79 | 200 | < .001 | 0.69 |
| Self-creation connection | 5.77 (1.15) | 4.45 (1.73) | 10.04 | 200 | < .001 | 0.71 |
| Intrinsic motivation | 4.93 (1.38) | 4.08 (1.66) | 7.09 | 200 | < .001 | 0.50 |
| Negative affect | 1.76 (1.11) | 1.85 (1.17) | −1.43 | 200 | .155 | −0.10 |
| *Guilt Item* | 1.48 (1.17) | 2.16 (1.64) | −5.99 | 200 | < .001 | −0.42 |
| *Shame Item* | 1.55 (1.31) | 1.99 (1.56) | −3.88 | 200 | < .001 | −0.27 |
| Positive affect | 5.49 (1.08) | 4.56 (1.55) | 8.24 | 200 | < .001 | 0.58 |
| Extrinsic motivation (expl) | 3.47 (1.85) | 3.16 (1.73) | 2.80 | 196 | .006 | 0.20 |
| Self-assessed creativity (expl) | 5.40 (1.29) | 4.48 (1.64) | 6.77 | 200 | < .001 | 0.48 |

*Note. Negative t and dz values indicate higher scores for AI-assisted work. Expl: Exploratory finding.*

*Work type & payment status.* To determine whether the differences between AI-assisted and human-only creations were driven by what participants created or whether they were paid for it, we compared the two conditions on work type and payment status. For both conditions, the most frequently chosen options were writing (human-only: 59, AI-assisted: 59), and visual art and design (human-only: 60, AI-assisted: 57; Tables S1-2), suggesting that type of work was overall comparable. The biggest difference was for craft, hobby, or DIY works (human-only: 20, AI-assisted: 4), though this represented a small percentage of responses. Payment status was also similar across conditions, with most participants reporting that they were not paid for their work (human-only: 149; AI-assisted: 142). Together, this suggests that the effects were driven primarily by creation mode rather than external factors.[2]

Study 1 provides initial evidence that AI threatens the perception that a creative work is "made in the image of its creator:" creators feel less authentic, less self-creation connection, less positive affect (and more guilt and shame), less intrinsically motivated, and rate their work as less creative. Nevertheless, Study 1 had several limitations: first, our findings offer little insight into the mechanisms underlying these effects; second, our sample was limited to U.S. citizens, a population which is particularly nervous and unexcited about AI (Stanford Institute for Human-Centered Artificial Intelligence, 2026); third, the negative PANAS items may not have been specific or sensitive enough to fully support or reject our “existential ick” hypothesis; fourth, the study was not pre-registered. We address these limitations in Study 2.

## 5. Study 2

[2] *Order effects.* Exploratory mixed-design ANOVA testing for each hypothesized dependent variable revealed no order effects (all Fs ≤ 1.16, all ps ≥ .282).

In Study 2, we sought to confirm the findings of Study 1 in a pre-registered study with a U.K. sample, refine the measurement of negative affect with specific subscales for disgust, guilt, and shame, and test the hypothesized role of love as a statistical mediator. Inspired by the *Handmade Effect* research (Fuchs et al., 2015), we measured creators' perceptions of love, quality, effort, uniqueness, and expensiveness of their own work, predicting that love would mediate the effect of AI-assistance on authenticity even when the four other mediators were accounted for. As in Study 1, the study used a counterbalanced two-condition (AI-assisted vs. human-only) within-subjects design. All hypotheses and analyses were pre-registered on the Open Science Framework.

*Exploratory:* We also tested whether the effects predicted in H1–H6a-e remained significant after controlling for perceived effort. We considered whether effort-based valuation—what has been variously conceptualized as effort justification, the effort heuristic, or the IKEA Effect (Aronson & Mills, 1959; Kruger et al., 2004; Norton et al., 2012)—could be a possible alternative explanation for our findings. Because AI-assistance often means less effort is required to produce a given work, the differences in our outcomes could be due to participants' devaluation of lower-effort creations rather than an effect specific to AI use.

## 5.1 Method

### *5.1.1 Participants*

The sample size was determined by funding availability. A sensitivity power analysis indicated that 250 participants would provide power of .92 to detect the hypothesized indirect effect, assuming $r = .33$ ($d = 0.70$) between creation mode and authenticity (from Study 1), $r = .36$ between creation mode and symbolic love (from Fuchs et al., 2015), and $r = .30$ between love and authenticity. Two hundred fifty U.K. adults ($M_{age} = 32.4$ years, $SD = 10.6$, range: 18–68; 148 male, 98 female, 3 non-binary/third gender, 1 preferred not to say) were recruited via Prolific and compensated at minimum wage. Participants were eligible if they were 18 years or older, fluent in English, U.K. citizens living in the United Kingdom, and used AI at least 2 times per week. To further ensure that participants had created works both with and without AI, we included a required confirmation question after informed consent (see supplemental materials).

### *5.1.2 Procedure*

The procedure was identical to Study 1 except with changes to the measures as follows.

### *5.1.3 Measures*

The following scales were repeated from Study 1: the state authenticity scale ($\alpha_{AI}$ = .86, $\alpha_{human}$ = .81), self-creation connection scale ($\alpha_{AI}$ = .96, $\alpha_{human}$ = .93), the 10 PANAS positive affect items ($\alpha_{AI}$ = .95, $\alpha_{human}$ = .88), the 6 intrinsic ($\alpha_{AI}$ = .80, $\alpha_{human}$ = .74) and 6 extrinsic ($\alpha_{AI}$ = .90, $\alpha_{human}$ = .90) motivation items, and the single self-assessed creativity item. The PANAS negative affect items were not included, but replaced by the disgust, guilt, and shame subscales described below. Unless otherwise noted, items were rated on seven-point scales (1 = strongly disagree; 7 = strongly agree).

*Disgust, guilt, and shame.* Disgust was measured with three items (e.g., "I felt disgusted") used by Kirk and Givi (2025) to assess disgust related to AI use ($\alpha_{AI}$ = .89, $\alpha_{human}$ = .77). Guilt (e.g., "I felt guilty"; $\alpha_{AI}$ = .91, $_{\alpha}$human = .76) and shame (e.g., "I felt ashamed"; $\alpha_{AI}$ = .82, $\alpha_{human}$ = .81) were measured with three items each, adapted from the State Shame and Guilt Scale (Marschall et al., 1994). Items were rated on a seven-point scale (1 = very slightly or not at all; 4 = moderately; 7 = extremely), and randomized among the positive affect items.

*Mediators.* Five mediators for each condition were measured with 12 items adapted from Fuchs et al. (2015): 3 love items (e.g., “It contains love”; $\alpha_{AI}$ = .96, $\alpha_{human}$ = .93); 2 uniqueness items (e.g., "It is unique"; $\alpha_{AI}$ = .90, $\alpha_{human}$ = .80); 2 quality items (e.g., "It is of good quality"; $\alpha_{AI}$ = .91, $\alpha_{human}$ = .87); 3 effort items (e.g., "It was time-intensive to create"; $\alpha_{AI}$ = .88, $\alpha_{human}$ = .79); and 2 expensiveness items (e.g., "It would be valuable if I sold it"; $\alpha_{AI}$ = .92, $\alpha_{human}$ = .94).

## *5.2 Results and Discussion*

### *5.2.1 Confirmatory Findings*

*Study 1 Replication Hypotheses (H1-H5).* As in Study 1, participants reported significantly lower authenticity, self-creation connection, positive affect, and intrinsic motivation, and significantly higher disgust, guilt, and shame, for their AI-assisted than human-only creative work (Table 2). It is worthwhile to note that the means on the moral-emotion subscales were low in both conditions (all *M*s < 2 on 7-point scales), indicating a significant but modest increase in negative affect rather than intense distress during AI use. Extrinsic motivation, for which no hypothesis was pre-registered, did not differ between creation modes. Confirming Study 1's exploratory finding, participants also rated their AI-assisted work as significantly less creative than their human-only work.

*Hypotheses 6a-e.* As predicted, participants perceived their AI-assisted (vs. human-only) work as less imbued with love, less effortful, less unique, lower in quality, and less expensive, with the largest difference for love.

**Table 2**

*Study 2: Descriptive Statistics, Paired t Tests, and Effect Sizes for All Measures*

| | Human-only *M* (*SD*) | AI-assisted *M* (*SD*) | *t* | *df* | *p* | *dz* |
|---|---|---|---|---|---|---|
| Authenticity | 6.21 (0.81) | 4.97 (1.31) | 13.93 | 249 | < .001 | 0.88 |
| Self-creation connection | 5.43 (1.29) | 4.15 (1.71) | 10.99 | 249 | < .001 | 0.70 |
| Intrinsic motivation | 4.81 (1.21) | 4.21 (1.44) | 6.46 | 245 | < .001 | 0.41 |
| Disgust | 1.29 (0.75) | 1.57 (1.07) | −4.98 | 249 | < .001 | −0.31 |
| Guilt | 1.41 (0.87) | 1.84 (1.30) | −5.49 | 249 | < .001 | −0.35 |
| Shame | 1.40 (0.87) | 1.66 (1.04) | −4.31 | 249 | < .001 | −0.27 |
| Positive affect | 5.47 (0.97) | 4.56 (1.44) | 10.48 | 249 | < .001 | 0.66 |
| Love | 5.60 (1.40) | 3.92 (2.01) | 13.05 | 249 | < .001 | 0.83 |
| Effort | 5.39 (1.20) | 3.86 (1.72) | 11.77 | 248 | < .001 | 0.75 |
| Uniqueness | 5.42 (1.49) | 4.21 (1.87) | 9.43 | 248 | < .001 | 0.60 |
| Quality | 5.38 (1.32) | 4.92 (1.52) | 4.38 | 249 | < .001 | 0.28 |
| Expensiveness | 3.19 (1.91) | 2.83 (1.89) | 2.83 | 249 | .005 | 0.18 |
| Extrinsic motivation (expl) | 2.96 (1.69) | 3.07 (1.70) | −1.20 | 245 | .230 | −0.08 |
| Self-assessed creativity (expl) | 5.36 (1.28) | 4.48 (1.60) | 7.86 | 249 | < .001 | 0.50 |

*Note.* Negative t and dz values indicate higher scores for AI-assisted work. Expl: Exploratory findings.

*Mediation (H7).* We tested mediation using the path-analytic approach for two-condition within-participant designs employing the MEMORE statistical package (v3.0; Montoya & Hayes, 2017). Because the design was within-subjects, the mediated effect of creation mode was estimated from difference scores (human-only minus AI-assisted), with mean-centered mediator averages included as covariates following Judd et al.'s (2001) approach. In the model with love as the sole mediator, the total effect of creation mode on authenticity was significant (Table 3). Creation mode strongly predicted perceived love, and the love difference in turn predicted the authenticity difference (Figure 1). The indirect effect through love was significant, accounting for approximately 55% of the total effect; the direct effect also remained significant, indicating partial mediation. When all five mediators were considered simultaneously, the indirect effect through love remained significant and was substantially stronger than any other mediator. Smaller but significant indirect effects also emerged through effort and quality, whereas the indirect effects through uniqueness and expensiveness were not significant. The direct effect of creation mode remained significant, and the full model accounted for 43% of the variance in the authenticity difference. Pairwise contrasts confirmed that the indirect effect through love was significantly larger than the indirect effect through each of the four rival mediators (Table 4).

**Figure 1**

*Study 2: Mediation Model of Creation Mode on Authenticity Through Love*

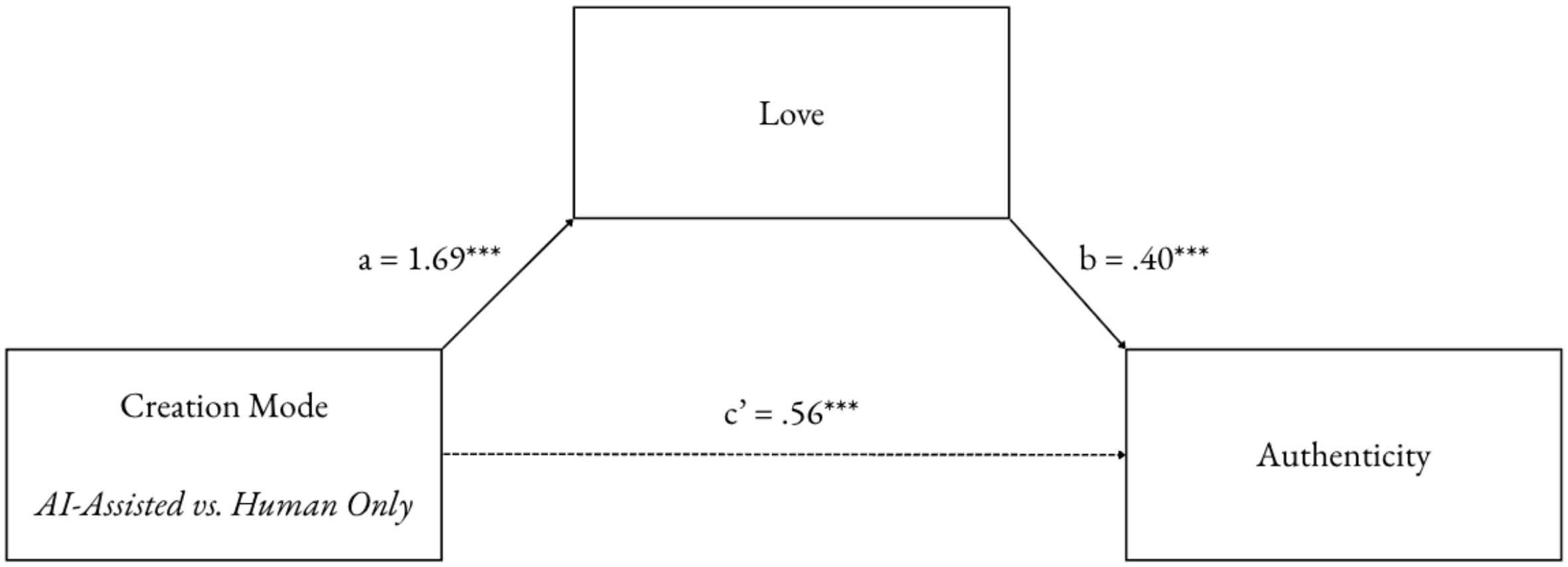

*Note.* ***$p < .001$.

**Table 3**

*Study 2: Effects of Condition on Authenticity Through Love and Rival Mediators*

| | Model 1 (Love only) *b* [95% *CI*] | Model 2 (All five) *b* [95% *CI*] |
|---|---|---|
| Indirect effect | | |
| Love | **0.68** [0.51, 0.87] | **0.51** [0.36, 0.70] |
| Effort | — | **0.16** [0.03, 0.31] |
| Quality | — | **0.05** [0.005, 0.12] |
| Uniqueness | — | 0.03 [−0.09, 0.14] |
| Expensiveness | — | −0.01 [−0.04, 0.02] |
| Direct effect | **0.56** [0.37, 0.75] | **0.50** [0.30, 0.71] |
| Total effect | **1.24** [1.07, 1.42] | **1.25** [1.07, 1.42] |

*Note.* Bootstrapped indirect effects (5,000 samples, bias-corrected). Model 1 N = 250; Model 2 N = 249. Bolded effects are significant at p < .05.

**Table 4**

*Study 2: Pairwise Contrasts Between the Indirect Effect of Love and Rival Mediators*

| | *b* | *95% CI* |
|---|---|---|
| Love vs. Effort | 0.35 | [0.15, 0.57] |
| Love vs. Uniqueness | 0.49 | [0.26, 0.74] |
| Love vs. Quality | 0.46 | [0.31, 0.64] |
| Love vs. Expensiveness | 0.52 | [0.36, 0.71] |

*Note.* All contrasts significant (95% CIs exclude zero), indicating the indirect effect through love was significantly stronger than through each rival mediator.

### *5.2.2 Exploratory Findings*

*Order effects.* Unlike in Study 1, repeated-measures ANOVA revealed significant order effects for six of the dependent variables (Table S3). The difference in self-creation connection, love, expensiveness, and positive affect scores (all lower for AI-assisted work) was larger when the human-only reflection came first; the difference in disgust and shame scores (both higher for AI-assisted work) was larger when the AI-assisted reflection came first. Authenticity, quality, effort, uniqueness, guilt, and intrinsic motivation showed no order effects.

*Controlling for effort.* A linear mixed model revealed that the human-only versus AI-assisted differences in authenticity, self-creation connection, love, uniqueness, and positive affect all remained significant after controlling for perceived effort (Table S4). In contrast, the intrinsic motivation and quality gaps disappeared after accounting for effort, suggesting these variables are sensitive to an effort heuristic rather than AI use specifically. Interestingly, the expensiveness gap remained significant but reversed direction, indicating that once effort is held constant, AI-assisted works are considered more expensive.

*Work type & payment status.* Once again, we compared the two conditions on work type and payment status. The most frequently chosen options were writing (human-only: 42, AI-assisted: 52), and visual art and design (human-only: 72, AI-assisted: 78; Tables S5-S6), suggesting that type of work was overall comparable. There was one notable outlier, however, with many more participants reporting craft, hobby, or DIY works in the human-only (59) than AI-assisted (25) condition. Payment status was similar across conditions, with most participants reporting that they were not paid for their work (human-only: 209; AI-assisted: 198).

*AI centrality and authenticity:* In an exploratory analysis, we tested whether the centrality of AI's role in the creative process—participants' self-reported rating of how central AI was to the work's creation (*1 = AI played little or no role, 7 = the work was created almost entirely by AI*)—correlated with authenticity scores in the AI-assisted condition. We found a small but significant relationship, such that greater AI centrality was associated with lower state authenticity ($r(248) = -.13$, $p = .038$). That this relationship is not larger, however, indicates that while "dose" matters to an extent, any degree of AI use seems to abate feelings of authenticity.

In summary, all Study 2 hypotheses were supported in a pre-registered design with a new demographic sample. The addition of our disgust, guilt, and shame subscales also confirmed Study 1's exploratory negative affect findings, indicating that AI-assistance gives rise to "existential ick" feelings, perhaps related to the sense that one is cheating, deceiving, or outsourcing an otherwise life-affirming act. Most importantly, we found that symbolic love—the perception that a creative work is imbued with a creator's passion and attachment—mediated over half of the authenticity gap between AI-assisted and human-only work, and did so significantly more strongly than the four rival mediators. This is consistent with our broader "made in the creator's image" theory: an AI-assisted work may be effortful, unique, or high quality, but if a creator does not feel that it is imbued with their love, it feels like a less authentic expression of themselves.

## 6. Study 3

In Study 3, we sought to replicate Study 2's mediation, as well as explore whether creators' feelings of love could be experimentally manipulated, and if this would in turn produce changes in authenticity. We used a 2 (creation mode: AI-assisted vs. human) × 2 (manipulation: love-up vs. love-down) design, in which all participants completed the standard AI-assisted condition, but were randomly assigned to a love-up or love-down manipulation. The manipulation only took place within the human condition, such that participants either completed the *AI-assisted, human: love-up* or *AI-assisted, human: love-down* condition.

## 6.1 Method

### *6.1.1 Participants*

Two hundred U.S. adults ($M_{age}$ = 39.4 years, *SD* = 13.7, range: 18–82; 105 male, 93 female, 2 non-binary/third gender) were recruited via Prolific and compensated at minimum wage. This sample size afforded us a power of 95% to detect effect sizes of *d* = .51 or higher, and met funding availability. Participant eligibility was the same as Study 1.

### *6.1.2 Procedure*

The procedure was identical to Studies 1 and 2 except as noted. Participants completed the two creative work reflection tasks (AI-assisted and human-only) in counterbalanced order, with prompt wording identical to Study 1. Following the human reflection prompt, participants were randomly assigned to the love-up or love-down manipulation. For the love-up manipulation, participants read a short passage that emphasized the passion, care, and emotional investment that goes into the creative process, and were prompted to consider that they likely felt passion and enjoyment while creating their human-only work; they then briefly described the parts of the work they felt particularly passionate about (prompts in supplemental materials). For the love-down manipulation, participants read a parallel passage emphasizing that the creative process can feel practical, routine, or unengaging, and were prompted to consider that they may not have felt particularly passionate about or invested in their human-only creation; they then briefly described the parts of the work they did not feel particularly passionate about. The language for the love manipulation was based on Fuchs et al. (2015) previously described conception of love. After each reflection task, participants completed the condition-specific

measures—the state authenticity scale and *Handmade Effect* measures—in randomized order, with respect to the work they had just described. Demographic information was collected at the end.

### *6.1.3 Measures*

The state authenticity scale ($\alpha_{AI}$ = .91, $\alpha_{human}$ = .86), and five mediators (love: $\alpha_{AI}$ = .97, $\alpha_{human}$ = .96; quality: $\alpha_{AI}$ = .92, $\alpha_{human}$ = .92; uniqueness: $\alpha_{AI}$ = .90, $\alpha_{human}$ = .87; expensiveness: $\alpha_{AI}$ = .91, $\alpha_{human}$ = .88, effort[3]) were measured.

## *6.2 Results and Discussion*

Our findings indicate that we were unable to successfully manipulate love—and perhaps therefore, authenticity (Table 5). Neither love nor authenticity differed significantly between love-up and love-down manipulations. Further, a repeated-measures ANOVA confirmed that while authenticity and love were significantly lower for AI-assisted than human-only work, in keeping with Studies 1 and 2, this gap did not depend on the love manipulation (Table 6). In short, our manipulation was not effective, reasons for which we explore in the general discussion. For the following analyses, we collapsed the love-up and love-down manipulation within the human-only condition, which allowed us to compare it to the AI-assisted condition, as in the prior studies.

**Table 5**

*Study 3: Descriptive Statistics, Independent t Tests, and Effect Sizes for Love-Up & Love-Down Manipulation*

| | Human-only: Love-up *M* (*SD*) | Human-only: Love-down *M* (*SD*) | *t* | *df* | *p* | *d* |
|---|---|---|---|---|---|---|

[3] Note. Cronbach's alpha is missing for effort. Due to a data collection error, effort Item 2 was excluded for the AI condition; effort was analyzed based on Item 1 for both conditions.

| | | | | | | |
|---|---|---|---|---|---|---|
| Love | 5.97 (1.31) | 5.64 (1.59) | 1.60 | 198 | .112 | 0.23 |
| Authenticity | 6.19 (0.98) | 6.10 (1.14) | 0.60 | 198 | .552 | 0.08 |

*Note.* Human-only: Love-up ($n$ = 99), human-only: Love-down ($n$ = 101).

**Table 6**

*Study 3: Repeated-Measures ANOVA*

| Within Subjects Effects | Sum of Squares | *df* | *F* | *p* | $n^2$ |
|---|---|---|---|---|---|
| Love | | | | | |
| Condition | 125.52 | 1, 198 | 63.45 | < .001 | 0.10 |
| Condition x Manipulation | 0.17 | 1, 198 | 0.09 | .77 | 0.00 |
| Authenticity | | | | | |
| Condition | 70.91 | 1, 198 | 56.87 | < .001 | 0.10 |
| Condition x Manipulation | 0.20 | 1, 198 | 0.16 | .69 | 0.00 |

**Table 7**

*Study 3: Descriptive Statistics, Paired t Tests, and Effect Sizes for All Measures*

| | Human-only *M* (*SD*) | AI-assisted *M* (*SD*) | *t* | *df* | *p* | *dz* |
|---|---|---|---|---|---|---|
| Authenticity | 6.15 (1.06) | 5.30 (1.45) | 7.55 | 199 | < .001 | 0.53 |
| Love | 5.81 (1.46) | 4.69 (1.92) | 7.98 | 199 | < .001 | 0.56 |
| Effort | 5.79 (1.45) | 4.33 (2.04) | 8.90 | 199 | < .001 | 0.63 |
| Quality | 5.79 (1.21) | 5.34 (1.53) | 4.08 | 199 | < .001 | 0.29 |
| Uniqueness | 5.77 (1.37) | 5.10 (1.70) | 5.12 | 199 | < .001 | 0.36 |
| Expensiveness | 4.31 (1.94) | 3.56 (1.99) | 5.46 | 199 | < .001 | 0.39 |

Running the same mediation analysis as in Study 2, we did, however, replicate Study 2's love mediation findings (Figure 2, Tables 7-9).

**Figure 2**

*Study 3: Mediation Model of Creation Mode on Authenticity Through Love*

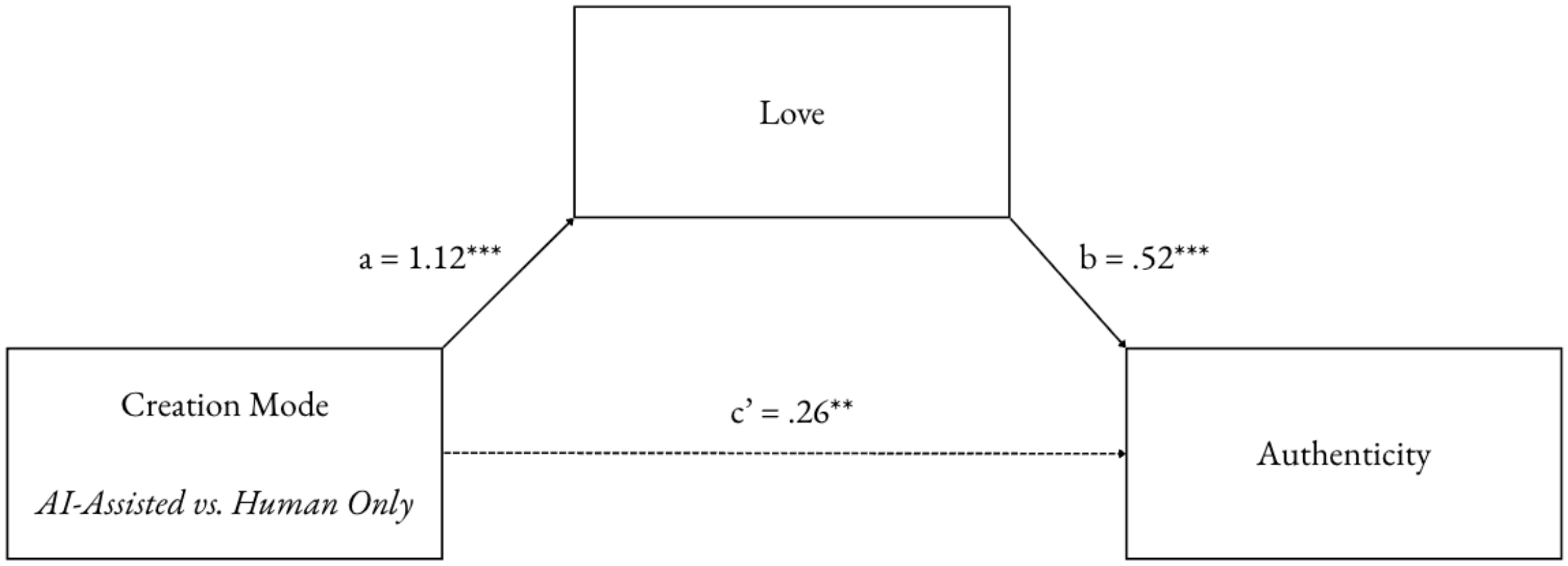

*Note.* $^{**}p < .01$, $^{***}p < .001$

**Table 8**

*Study 3: Effects of Condition on Authenticity Through Love and Rival Mediators*

| | Model 1 (Love only) *b* [95% *CI*] | Model 2 (All five) *b* [95% *CI*] |
|---|---|---|
| Indirect effect | | |
| Love | **0.58** [0.41, 0.79] | **0.44** [0.26, 0.66] |
| Effort | — | 0.09 [−0.06, 0.25] |
| Quality | — | 0.03 [−0.05, 0.15] |
| Uniqueness | — | 0.04 [−0.07, 0.17] |
| Expensiveness | — | 0.02 [−0.09, 0.13] |
| Direct effect | **0.26** [0.07, 0.45] | **0.22** [0.01, 0.43] |
| Total effect | **0.84** [0.62, 1.06] | **0.84** [0.62, 1.06] |

*Note.* Bootstrapped indirect effects (5,000 samples, bias-corrected). $N = 200$. Bolded effects are significant at $p < .05$.

**Table 9**

*Study 3: Pairwise Contrasts Between the Indirect Effect of Love and Rival Mediators*

| | *b* | *95% CI* |
|---|---|---|
| Love vs. Effort | 0.36 | [0.12, 0.61] |
| Love vs. Uniqueness | 0.40 | [0.16, 0.69] |
| Love vs. Quality | 0.41 | [0.19, 0.67] |
| Love vs. Expensiveness | 0.42 | [0.23, 0.68] |

*Note.* All contrasts significant (95% CIs exclude zero), indicating the indirect effect through love was significantly stronger than through each rival mediator.

## 7. General Discussion

Across three studies, we found that AI-assisted creative work is a less effective means of expressing the self, one which appears to compromise myriad psychological boons typically associated with creativity. Relative to their own human-only work, creators report lower state authenticity, weaker self-creation connection, flattened positive affect (what we term a "creative anhedonia" effect), and reduced intrinsic motivation when reflecting on work made with AI, as well as modest but significant increases in the "existential ick" emotions of disgust, guilt, and shame. Revealingly, creators also rated their AI-assisted works as significantly less creative. These effects replicated across three independent samples from two countries (U.S. and U.K.), and did not appear to be dependent on external factors, such as work type or payment status. Further, this authenticity gap was statistically mediated by the perception that AI-assisted works contain less of the creator's love, above and beyond the four rival mediators. With the exception of intrinsic motivation, expensiveness, and quality, these effects remained significant after controlling for effort, abating concerns that creators are simply more partial to work that they have put more effort into.

Taken together, these findings extend a growing body of research on how audiences evaluate AI usage in creative work (e.g., Kirk & Givi, 2025; Millet et al., 2023; Raj et al., 2026) by turning the attention for the first time, to the best of our knowledge, to the creator's internal experience of self and identity. We provide initial evidence that AI use in the creative process interrupts the expression of self, and gives rise to the feeling in creators that the work is not made in their own image. As a result, the downstream benefits long associated with creativity—a sense of authentic authorship, positive affect, intrinsic motivation, and appraising one's own creativity favorably—attenuate, and adverse outcomes arrive in their stead. We propose that these effects can be largely explained by self-determination theory, in which authenticity is seen as contingent on autonomous, intrinsically-driven engagement (Goldman & Kernis, 2002; Ryan & Deci, 2004). In AI-assisted work, however, the creator's role shifts to something closer to a curator, which may minimize the sense that they are indeed self-determining the outcome. We are also the first, to our knowledge, to extend the *Handmade Effect* to AI, by replacing machine-made works with AI-assisted works, and situating the inquiry from the perspective of the creator rather than the consumer. While Fuchs and colleagues (2015) and the essentialism and contagion literature (Newman, 2016; Newman et al., 2011) are concerned with the consumer's perception of a work's essence or love, we concern ourselves with the creator. That love not only mediated the authenticity gap, but did so even when accounting for the four rival mediators, suggests that feelings of authenticity are bound up in a creator's passionate attachment to their creation—or more poetically, that it is in putting their love into what they create that they come to see themselves more fully in it.

These findings have meaningful implications for creators, audiences, and cultures alike. Despite the protests of many, our world is rapidly converging around a future with AI at its center (Corazza et al., 2025). Work across corporate, scientific, academic, and creative settings alike is increasingly expected to be produced in collaboration with AI in the name of efficiency, quality, and cost savings. To be sure, AI is not without humanistic promise: it may replace mundane, repetitive work not worthy of a human's energy; it may democratize education and entrepreneurship in places where these are sorely needed; it may bring about medical or other innovative breakthroughs with curative potential. However, AI use that is extensive, passive, or conducted for tasks with high self-relevance—as is the case for much if not all creative work—appears to bring about a host of unwanted outcomes. From the perspective of the creator, we now have evidence that the self-expressive and self-constitutive functions of creativity are impinged upon in AI use, and that creators are less connected to and able to see themselves in their creation as a result. For audiences, this lack of humanity appears to yield, at best, perceptions that a work is less creative and less awe-inspiring (Millet et al., 2023), and at worst, moral hostility towards its artificial origins (Kirk & Givi, 2025; Koning & Voorveld, 2025; Schilke & Reimann, 2026). As for culture, it follows that a world in which more content is synthetically generated will produce a culture that is more homogenized, more derivative, and, both literally and figuratively, less human. More optimistically, it also follows that human-generated works may become all the more valued as they become more scarce.

However, several limitations apply across all three studies. First and perhaps most notably, our design relies on participants' reflection on creative works they produced in the past, rather than experimentally assigning participants to a creation mode. This was intentional: we believed that salient, meaningful feelings of authenticity, love, and the like would not be reliably elicited in an experimental environment. The creative process often unfolds slowly and over time, and we doubted such self-relevant feelings would be sufficiently triggered in response to an experimental task in which a participant may not feel invested nor passionate. Second, all measures relied on self-report. While this is appropriate for authenticity and our other self/identity constructs, which are definitionally subjective, it is also possible that responses in the AI-assisted condition were influenced by a sort of "AI creativity bias," rather than fully accurate self-reporting. To this end, our second study found order effects in which several positive outcomes (self-creation connection, love, expensiveness, and positive affect) showed a larger AI-assisted penalty when the human-only condition was presented first, and several negative outcomes (disgust and shame) showed a larger AI-assisted penalty when the AI-assisted condition was presented first. Third and relatedly, our participants were from the U.S. (Studies 1 and 3) and U.K. (Study 2), both countries with AI-skeptical populations (Stanford Institute for Human-Centered Artificial Intelligence, 2026). It is unclear if these effects would replicate, attenuate, or reverse in countries with more positive feelings and less negative feelings about AI, like China or Indonesia.

In Study 3, we were unable to successfully manipulate love. One explanation for our null findings is that our manipulation prompts were ineffective. This seems unlikely, however, as participants' responses generally indicated that they were able to successfully reflect on the passionate—or dispassionate—parts of their human-only work (sample responses in supplemental materials). A second and more interesting explanation, however, is that—when it comes to human-made creative work—feelings of love may be highly persistent. Once a creator has put their love into a work by way of an intimate, one-to-one relationship with it, those love feelings may withstand experimental prodding towards a less romantic conclusion. In other words, love may persist—and perhaps in keeping, feelings of authenticity remain unswayed. It is worthwhile to note that authenticity in the human condition—and to a lesser degree, love—showed ceiling effects, with authenticity's love-up ($M_{auth} = 6.19$) and love-down ($M_{auth}=6.10$) means nearing the ceiling of 7. This was not the case with AI-assisted work ($M_{auth} = 5.30$, $M_{love} = 4.69$). To this end—and potentially explaining our inability to manipulate these measures—a "halo effect" (Nisbett & Wilson, 1977; Noor et al., 2023) may apply to human creativity whereby participants are less able or willing to see their self-made creations as anything but love-filled and authentic. Conversely, as explored, research demonstrates that AI creative work suffers from a bias (Horton et al., 2023; Millet et al., 2023)—which in broader literature may be considered a "horn effect" (Noor et al., 2023)—whereby creators and consumers judge it harshly in ways that often defy the rational.

We hope this research sparks further work in related inquiries. Of greatest present interest to us is whether a creator's perception of their own authenticity and love in human vs. AI-assisted work can be detected by audiences. In other words, when a creator feels as though a work is—or is not—made in their own image, do audiences feel the same? This would indicate that authenticity is not merely tied to *beliefs* about a work which are ultimately violable (e.g., beliefs about human vs. artificial origins), but something *inherent* to the work itself (e.g., vulnerability, imperfection, stylistic idiosyncrasies, etc.). Such a finding would bolster claims that creativity produced by AI—while capable of satisfying the SDC—is not equivalent to human creativity insofar as it lacks authenticity. Another avenue to explore is how "dose-responsive" the effects of AI use on authenticity and love are. Our exploratory analyses found that, in the AI-assisted condition, the more central a role AI played in the creative process, the less authentic the creator felt. This modest but significant finding suggests more work needs to be done to manipulate the degree of AI involvement to determine if there is a minimal threshold at which a creator's self-expression is preserved and a maximum threshold at which it disappears. Additionally, given findings that creative expertise moderates certain adverse outcomes of AI use—specifically, that experts' ownership and agency take a greater hit than those of novices (He, 2026)—it is relevant to examine whether factors like domain expertise or creative identity/self-concept moderate the authenticity and love findings. Finally, as noted prior, it is important to replicate these findings in countries that are not WEIRD (Western, Educated, Industrialized, Rich, Democratic; Henrich et al., 2010).

We conclude simply with the statement that, relative to human-only work, creators are less able to see their AI-assisted creative works as "made in their image." This is a somber thought, as it interrupts two fundamental functions of creativity: that it conveys the authentic self of its creator, and that this self can be known by, resonate with, and find belonging in others. As our world converges around an AI future, it will be essential to maintain the humanity that has long made creativity a source of individual and collective thriving. This sentiment is echoed in the words of filmmaker Guillermo del Toro (2025) when asked about AI:

"I think [the artist's] duty is to share experience... This is how it feels to be me, from my point of view. When we speak of ourselves, we're not only doing self-portraiture. We're also creating a mirror for other people to say: this pain I hear in this song, this loss I hear in this poem, is similar to mine. And then in that moment you belong. You belong to humanity."